\documentclass[twocolumn,preprintnumbers,amsmath,amssymb]{revtex4}

\usepackage{graphicx}
\usepackage{dcolumn}
\usepackage{bm}

\begin{document}

\preprint{}

\title{A generative model for feedback networks}

\author{Douglas R. White}
 \altaffiliation[Also at ]{Santa Fe Institute, 1399 Hyde Park Road, Santa Fe, NM 87501, USA}
 \affiliation{%
 Institute of Mathematical Behavioral Sciences, University of California Irvine, Irvine, CA 92697, USA
 }%

\author{Nata\v sa Kej\v zar}
 \altaffiliation[Also at ]{Santa Fe Institute, 1399 Hyde Park Road, Santa Fe, NM 87501, USA}
 \affiliation{%
 Faculty of Social Sciences, University of Ljubljana, Kardeljeva pl. 5, 1000 Ljubljana, Slovenia
 }%

\author{Constantino Tsallis}
 \altaffiliation[Also at ]{Santa Fe Institute, 1399 Hyde Park Road, Santa Fe, NM 87501, USA}
 \affiliation{%
 Centro Brasileiro de Pesquisas F\'{\i}sicas, Xavier Sigaud 150, 22290-180 Rio de Janeiro-RJ, Brazil
 }%

\author{Doyne Farmer}
 \affiliation{%
 Santa Fe Institute, 1399 Hyde Park Road, Santa Fe, NM 87501, USA
 }%

\author{Scott White}
 \affiliation{%
 School of Information and Computer Science, University of California Irvine, Irvine, CA 92697, USA
 }%




\date{\today}

\begin{abstract}
We investigate a simple generative model for network formation. The model is designed to describe the growth of networks of kinship, trading, corporate alliances, or autocatalytic chemical reactions, where feedback is an essential element of network growth.  The underlying graphs in these situations grow via a competition between cycle formation and node addition.  After choosing a given node, a search is made for another node at a suitable distance.  If such a node is found, a link is added connecting this to the original node, and increasing the number of cycles in the graph; if such a node cannot be found, a new node is added, which is linked to the original node.  We simulate this algorithm and find that we cannot reject the hypothesis that the empirical degree distribution is a $q$-exponential function, which has been used to model long-range processes in nonequilibrium statistical mechanics.
\end{abstract}


\maketitle

\section{INTRODUCTION}

We present a generative model for constructing networks that grow via competition between cycle formation and the addition of new nodes.  The algorithm is intended to model situations such as trading networks, kinship relationships, or business alliances, where networks evolve by either establishing closer connections by adding links to existing nodes or alternatively by adding new nodes.  In arranging a marriage, for example, parents may attempt to find a partner within their pre-existing kinship network.  For reasons such as alliance building and incest avoidance, such a partner should ideally be separated by a given distance in the kinship network \cite{White Houseman}.  Such a marriage establishes a direct tie between families, creating new cycles in the kinship network.  Alternatively, if they do not find an appropriate partner within the existing network, they may seek a partner completely outside it, thereby adding a new node and expanding it.

Another motivating example is trading networks \cite{White Spufford}.  Suppose two agents (nodes) are linked if they trade directly.  To avoid the markups of middlemen, and for reasons of trust or reliability, an agent may seek new, more distant, trading partners.  If such a partner is found within the existing network a direct link is established, creating a cycle.  If not, a new partner is found outside the network, a direct link is established, and the network grows.   A similar story can be told about strategic alliances of businesses \cite{Jain, White Powell}; when a business seeks a partner, that partner should not be too similar to businesses with which relationships already exist.  Thus the business will first take the path of least effort, and seek an appropriate partner within the existing network of businesses that it knows;  if this is not possible, it may be forced to find a partner outside the existing network.

All of these examples share the common property that they involve a competition between a process for creating new cycles within the existing network and the addition of new nodes to the network.  While there has been an explosion of work on generative models of graphs \cite{Albert Barabasi, Bollobas Riordan, Adamic,Soares,Thurner}, there has been very little work on networks of this type.  The only exception that we are aware of involves network models of autocatalytic metabolisms \cite{Jain,Kauffman,Rossler,Farmer,Bagley}.  Such autocatalytic networks have the property that network growth comes about through the addition of autocatalytic cycles, which can either involve existing chemical species or entirely new chemical species.   Previous work has focused on topological graph closure properties \cite{Kauffman, Farmer}, or the simulation of chemical kinetics \cite{Bagley}, and was not focused on the statistical properties of the graphs themselves.  We call graphs of the type that we study here {\it feedback networks} because the cycles in the graph represent a potential for feedback processes, such as strengthening the ties of an alliance or chemical feedback that may enhance the concentration corresponding to an existing node \cite{White Houseman}.

We study the degree distributions of the graphs generated by our algorithm \cite{Albert Barabasi, Bollobas Riordan, Bollobas Palmer}, and find that they are well-described by distribution functions that have recently been proposed in nonequilibrium statistical mechanics, more precisely in nonextensive statistical mechanics  \cite{Tsallis, Gell-Mann Tsallis}.   Such distributions occur in the presence of strong correlations, e.g. phenomena with long-range interactions. Our intuition for why these distributions occur here is that the cycle generation inherently generates long range correlations in the construction of the graph.

\section {MODEL}

The growth model we propose closely mimics the examples given above.  For each time step, a starting node $i$ is randomly selected (e.g. the person or family looking for a marriage partner) and a target node $j$ (the marriage partner) is searched for within the existing network.  Node $j$ is not known at the outset but is searched for starting at node $i$. The search proceeds by attempting to move through the existing network for some $d$ number of steps without retracing the path.  If the search is successful a new link (edge) is drawn from $i$ to $j$.  If the search is unsuccessful, as explained below, a new node $j^\prime$ is added to the graph and a link is drawn from $i$ to $j^\prime$.  This process can be repeated for an arbitrary number of steps. In our simulations, we begin with a single isolated node but the initial condition is asymptotically not important. 

For each time step we randomly draw from a scale free distribution the starting node $i$, the distance $d$ (number of steps necessary to locate $j$ starting at $i$ assuming that such a location does occur), and for each node along the search path, the subsequent neighbor from which to continue the search. While node $j$ isn't randomly selected at the outset, it is obviously guaranteed that the shortest path distance from $i$ to $j$ is at most $d$. 
We now describe the model in more detail including the method for generating search paths, and the criterion for a successful search.

\begin{itemize}
\item \textbf{\textit{ Selection of node $i$}}.  The probability $\mbox {P}_{\alpha }$ of selecting a given node from among the $N$ nodes of the existing network is proportional to its degree raised to a power $\alpha$.  The parameter $\alpha > 0$ is called the \textit{attachment parameter}.

\begin{equation}
   \mbox {P}_\alpha (i) = {{\mbox [{deg}(i)]^{\alpha}} \over {{\sum_{m=1}^{N}} {\mbox [{deg}}(m)]^{\alpha}}}
   \label{eq1}
\end{equation}

\item \textbf{\textit{Assignment of search distance $d$}}.  An integer $d$ is chosen with probability $P_{\beta}$ where $\beta > 1$ is the \textit{distance decay parameter}. \footnote{ $\beta > 1$ is required to make the sum in the normalization converge.}.

\begin{equation}
   \mbox {P}_\beta (d) = {{d^{-\beta}} \over {{\sum_{m=1}^{\infty}} {m^{-\beta}}}}
   \label{eq2}
\end{equation}

\noindent In our experiments, we use the approximation of $\sum_{m=1}^{10^5} m^{-\beta}$ for computing the denominator of Eq. \ref{eq2}.

\item \textbf{\textit{Generation of search path}}. In the search for node $j$, assume that at a given instant  the search is at node $r$, where initially $r = i$.  A step of the search occurs by randomly choosing a neighbor of $r$, defined as a node $l$ with an edge connecting it to $r$.  We do not allow the search to retrace its steps, so nodes $l$ that have already been visited are excluded.  Furthermore, to make the search more efficient, the probability of choosing node $l$ is weighted based on its unused degree $u(l)$, which is defined as the number of neighbors of $l$ that have not yet been visited \footnote{By this we mean the number of neighboring nodes that have not been visited on this step of the algorithm, i.e. on this particular search.}.  The probability for selecting a given neighbor $l$ is
\begin{equation}
   \mbox {P}_\gamma (l) = {{\Big[1+ {\mbox u}(l)^\gamma\Big]} \over {\sum_{m=1}^M {\Big[1+{\mbox u}(m)^\gamma\Big]}}},
   \label{eq3}
\end{equation}
where $M$ is the number of unvisited nearest neighbors of node $r$.  $\gamma > 0$ is called the {\it routing parameter}. If there are no unvisited neighbors of $r$ the search is terminated, a new node is created, and an edge is drawn between the new node and node $i$.  Otherwise this process is repeated up to $d$ steps, and a new edge is drawn between node $j = l$ and node $i$.  In the first case we call this {\it node creation}, and in the second case, {\it cycle formation}.
\end{itemize}

\section{RESULTS}

\begin{figure*}[hbp]
\begin{center}
\setlength{\unitlength}{0.6pt}
  \begin{picture}(700,700)
   \put(0,0){\includegraphics[width=70mm,bb= 75 20 1035 900,clip=]{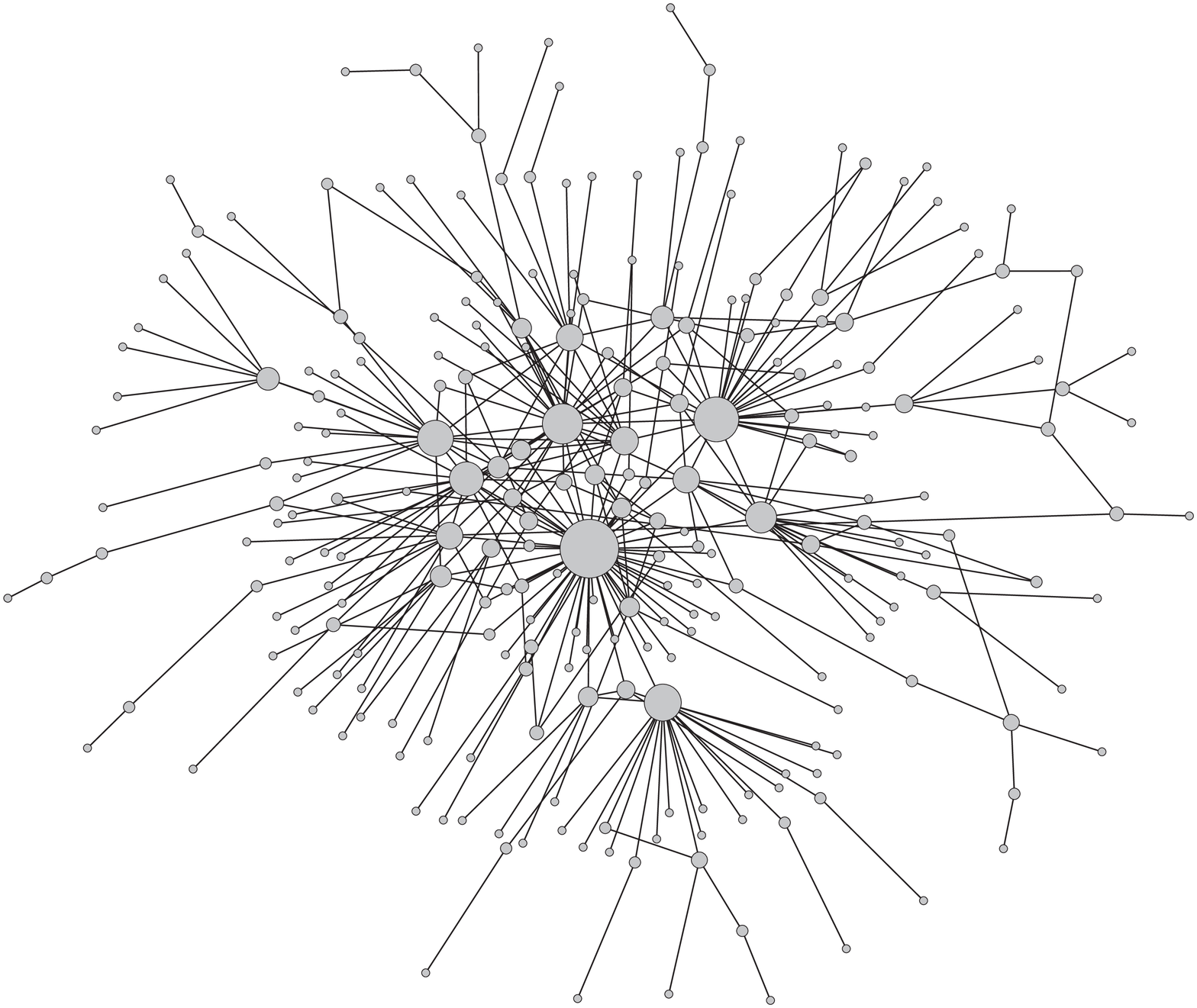}}
   \put(0,0){\small (c)}
   \put(350,0){\includegraphics[width=70mm,bb= 75 20 1035 900,clip=]{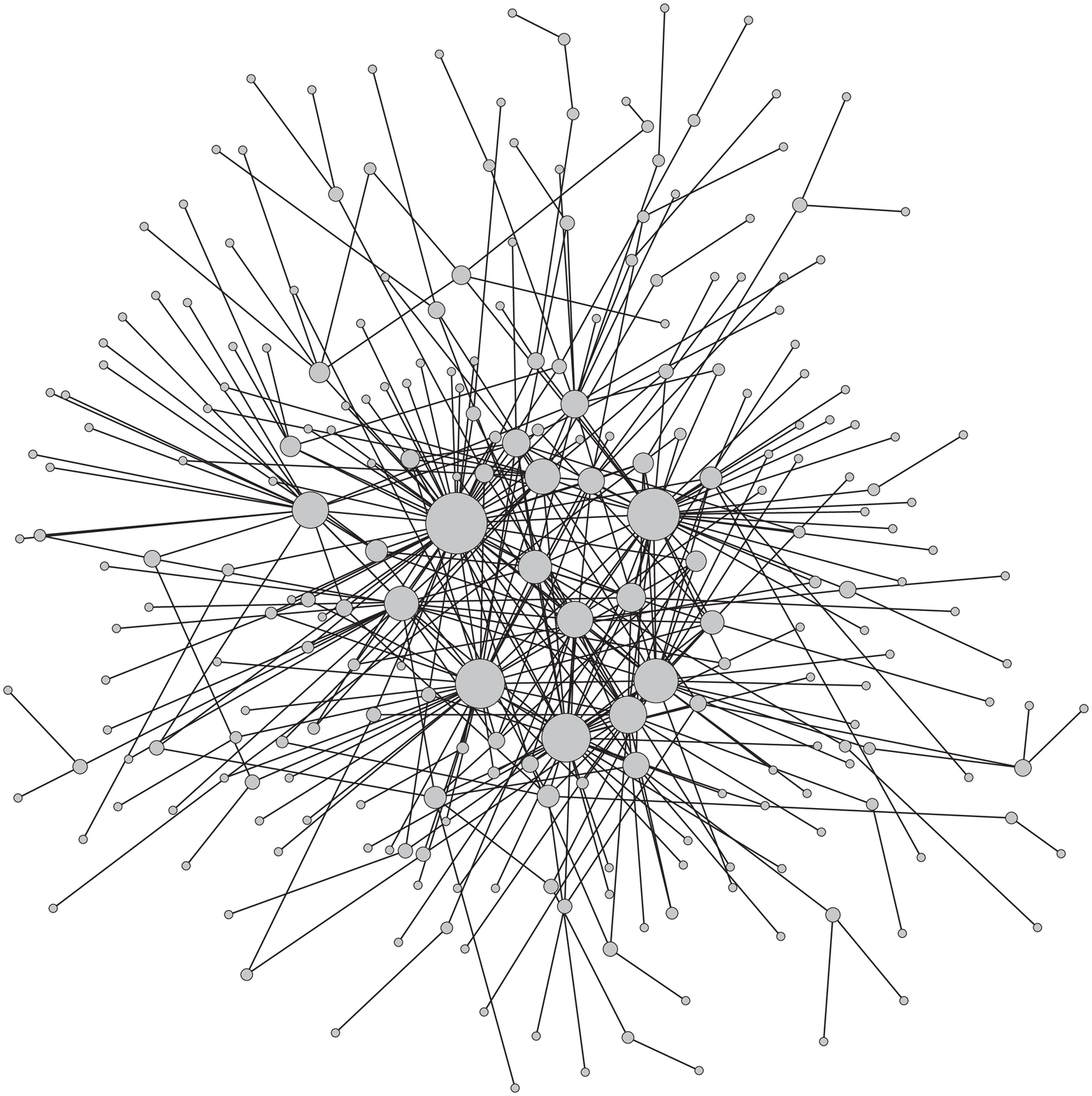}}
   \put(350,0){\small (d)}
   \put(0,350){\includegraphics[width=70mm,bb= 75 20 1035 900,clip=]{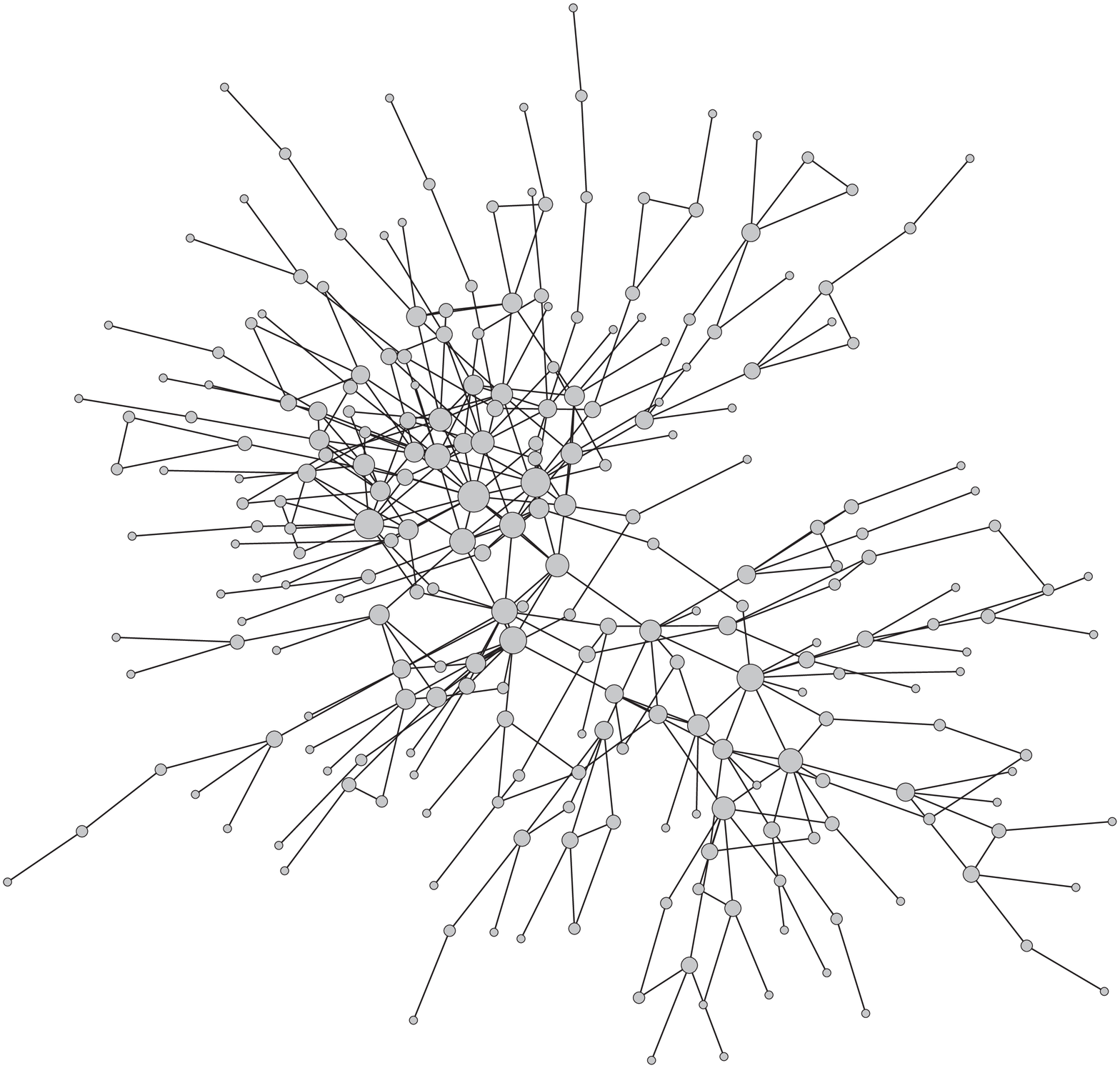}}
   \put(0,350){\small (a)}
   \put(350,350){\includegraphics[width=70mm,bb= 75 20 1035 900,clip=]{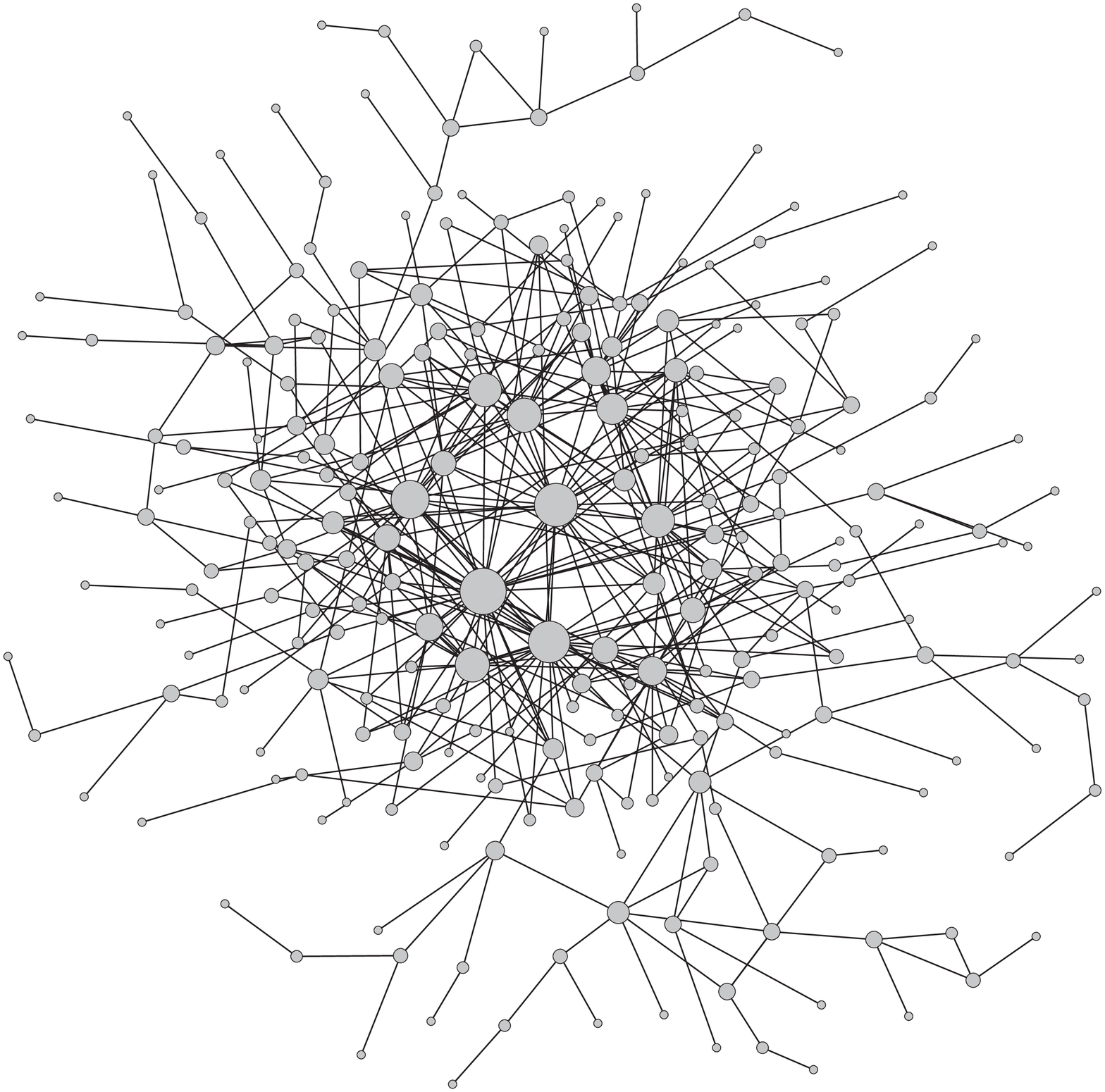}}
   \put(350,350){\small (b)}
  \end{picture}
\end{center}
\caption{Representations of typical network models with 250 nodes for $\beta = 1.3$. The panels correspond to (a) $\alpha = 0, \gamma = 0$, (b) $\alpha = 0, \gamma = 1$, (c) $\alpha = 1, \gamma = 0$ and (d) $\alpha = 1, \gamma = 1$. Sizes of nodes are proportional to their degrees. In the bottom graphs hubs emerge spontaneously due to preferential attachment ($\alpha  = 1$) while on the right more clustering occurs because of the larger routing parameter in cycle formation ($\gamma  = 1$). Notice that the denomination {\it preferential attachment} is also used in the literature in a slightly different sense, namely when the probability of a new node to attach to a pre-existing one of the growing network is proportional to the degree of the pre-existing one.}
\label{degree graphs}
\end{figure*}

\begin{figure*}[htp]
\begin{center}
\setlength{\unitlength}{0.6pt}
  \begin{picture}(700,700)
   \put(0,0){\includegraphics[width=70mm,bb= 75 20 1035 900,clip=]{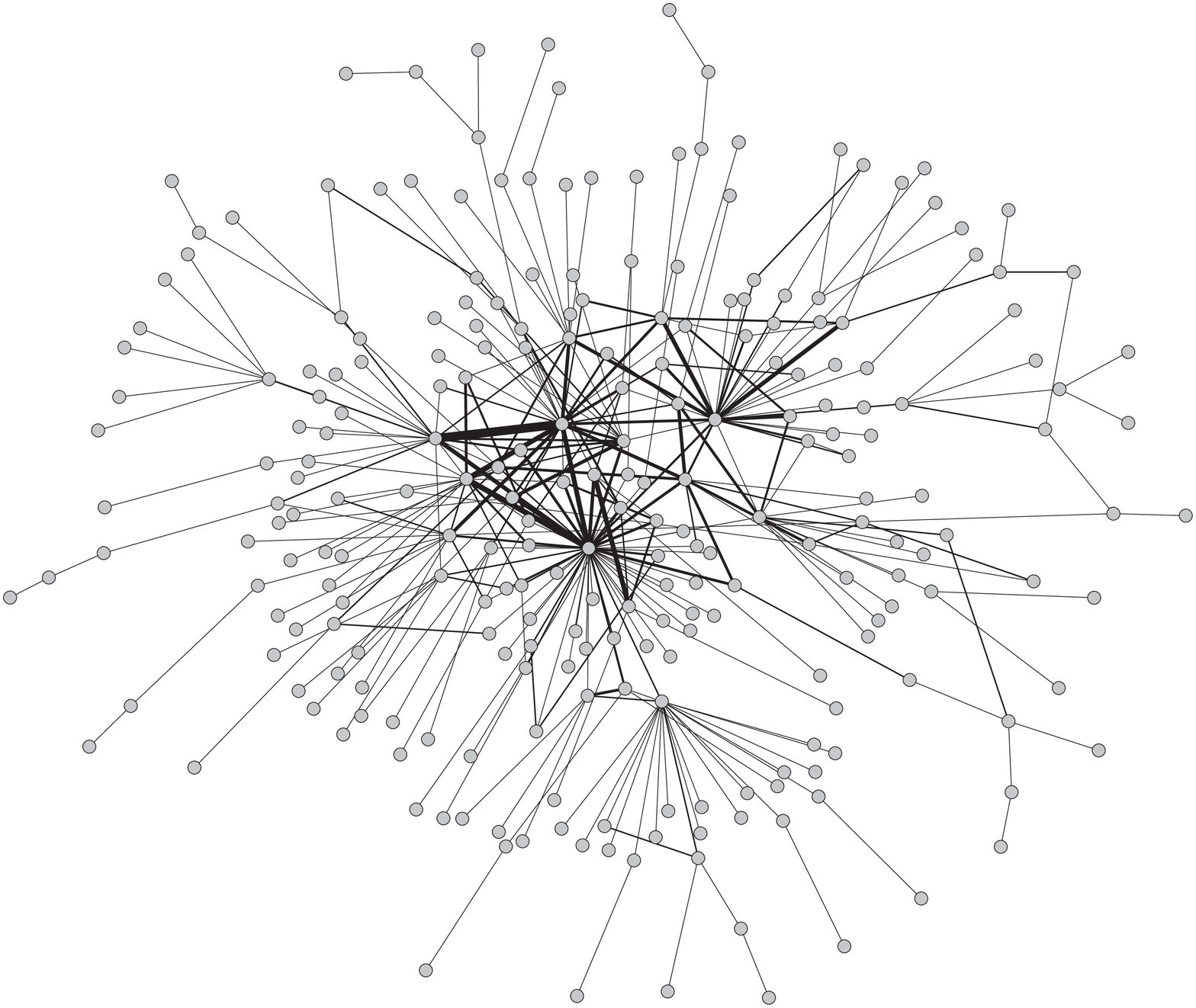}}
   \put(0,10){\small (c)}
   \put(350,0){\includegraphics[width=70mm,bb= 75 20 1035 900,clip=]{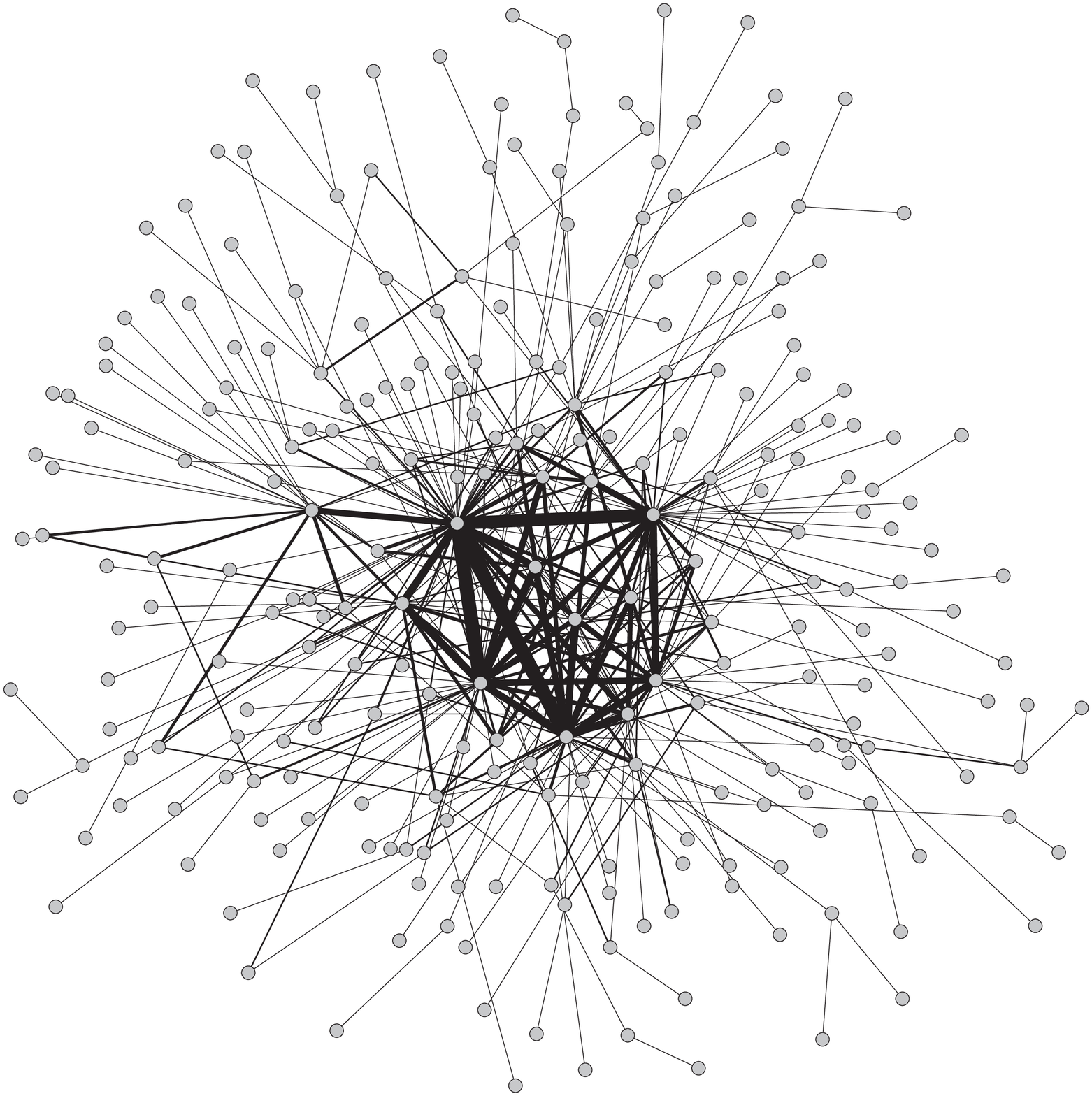}}
   \put(350,10){\small (d)}
   \put(0,350){\includegraphics[width=70mm,bb= 75 20 1035 900,clip=]{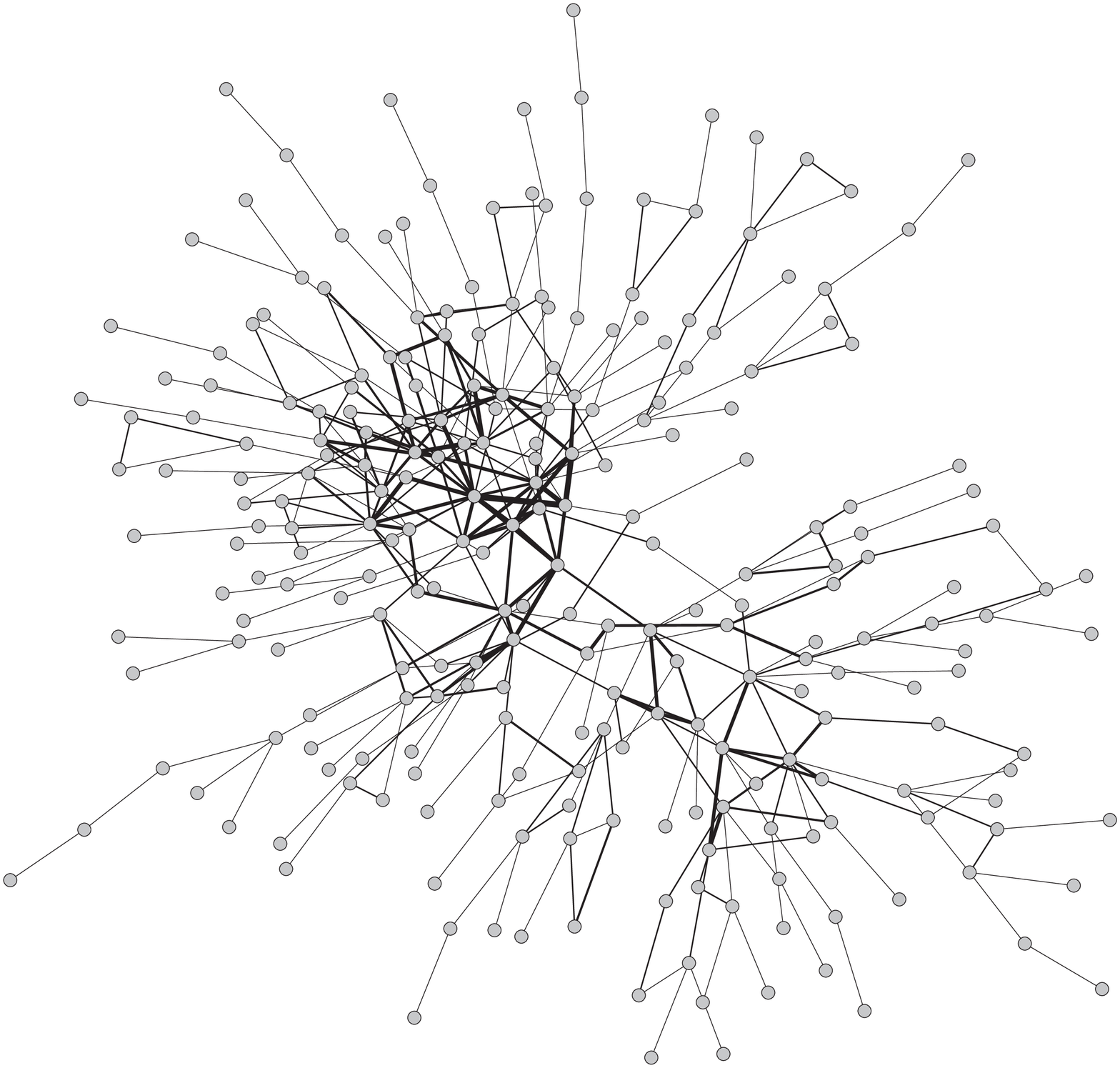}}
   \put(0,360){\small (a)}
   \put(350,350){\includegraphics[width=70mm,bb= 75 20 1035 900,clip=]{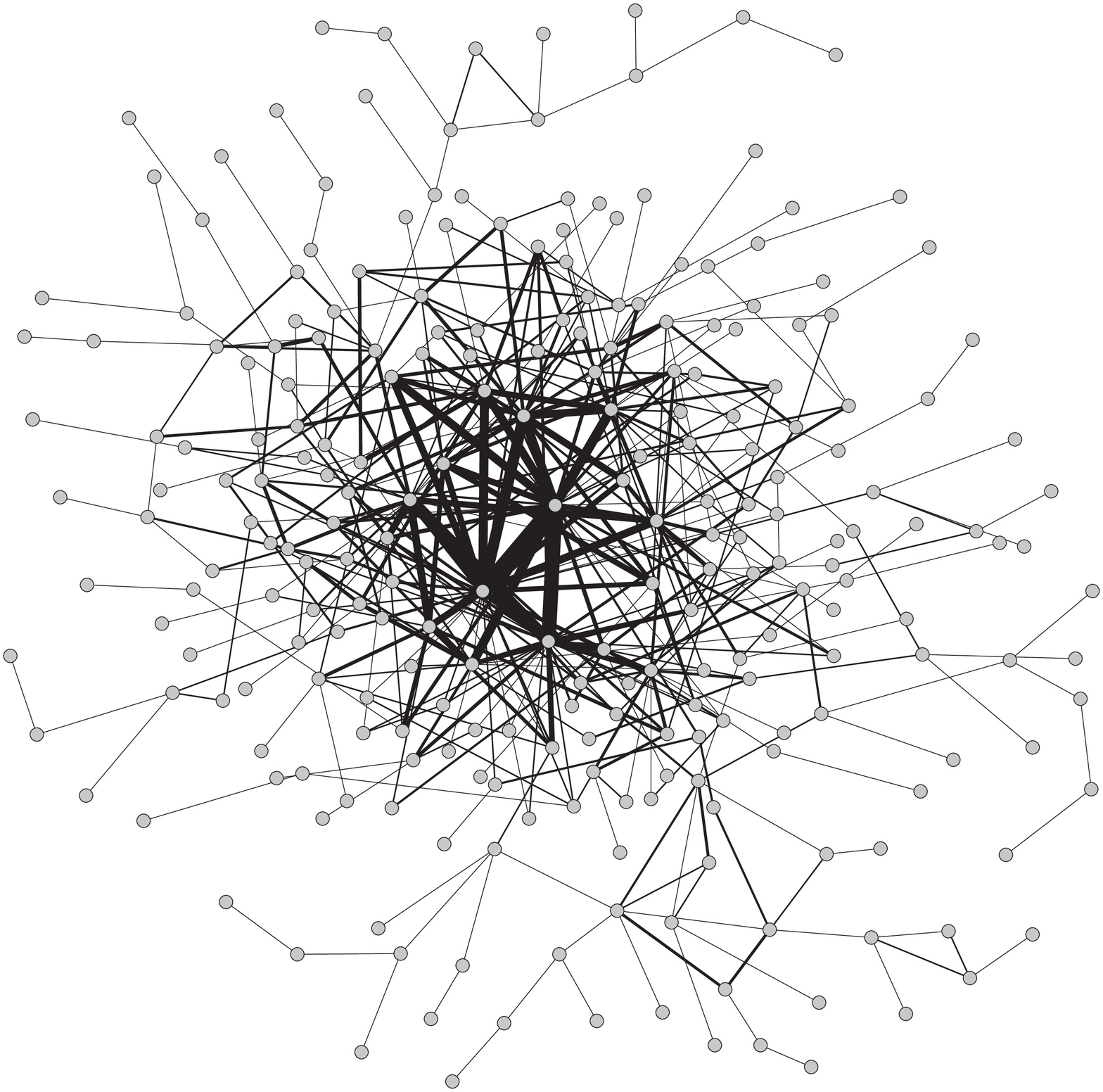}}
   \put(350,360){\small (b)}
  \end{picture}
\end{center}
\caption{Representations of typical network models with 250 nodes for $\beta = 1.3$. The panels correspond to (a) $\alpha = 0, \gamma = 0$, (b) $\alpha = 0, \gamma = 1$, (c) $\alpha = 1, \gamma = 0$ and (d) $\alpha = 1, \gamma = 1$. The thickness of an edge is proportional to the number of successfully created feedback cycles in which the edge has participated. The networks on the right of Figs. \ref{degree graphs} and \ref{weight graphs} show clusters of connected hubs with well-traversed routes around the clusters, while in those on the left, more tree-like, hubs connect but not in clusters with well-traversed routes around them.}
\label{weight graphs}
\end{figure*}

Typical feedback networks with $N=250$ for $(\alpha,\beta, \gamma)$ of $\{(0,1.3,0)$, $(0,1.3,1)$, $(1,1.3,0)$, $(1,1.3,1)\}$ are shown in Figures \ref{degree graphs} and \ref{weight graphs}.
The two figures display different depictions of the same four graphs. In Figure \ref{degree graphs} the sizes of the nodes represent their degrees and in Figure \ref{weight graphs} the thickness of the edge is proportional to the number of successfully created feedback cycles in which the edge participated (i.e. the number of times the search traversed this edge).

The attachment parameter $\alpha$ controls the extent to which the graph tends to form hubs (highly connected nodes).  When $\alpha = 0$ there is no tendency to form hubs, whereas when $\alpha$ is large there tend to be fewer hubs.
As the distance decay parameter $\beta$ increases the network tends to become denser due to the fact that $d$ is typically very small.   As $\gamma$ increases the search tends to seek out nodes with higher connectivity, there is a higher probability of successful cycle formation, and the resulting graphs tend to be more interconnected and less tree-like.

Despite that fact that network formation in our model depends purely on local information, i.e. each step only depends on information about nodes and their nearest neighbors, the probability of cycle formation is strongly dependent on the global properties of the graph, which evolve as the network is being constructed. 
In our model there is a competition between successful searches, which increase the degree of two nodes and leave the number of nodes unaltered, and unsuccessful searches, which increase the degree of an existing node but also create a new node with degree one.  Successful searches lower the mean distance of a node to other nodes, and failed searches increase this distance.  This has a stabilizing effect -- a nonzero rate of failed searches is needed to increase distances so that future searches can succeed. Using this mechanism to grow the network ensures that local connectivity structures, in terms of the mean distance of a node to other nodes, are somewhat similar across nodes thus creating long-range correlations between nodes.
Because these involve long-range interactions, we check whether the resulting degree distributions can be described by the form
\begin{equation}
   p(k) = p_0 k^{\delta} e_{q}^{- {k/\kappa}}
   \label{eq4}
\end{equation}
\noindent where the {\it $q$-exponential} function $e_q^x$ is defined as
\begin{equation}
   e_{q}^{x} \equiv \Big[1 + (1 - q) x\Big]^{{1}/{(1-q)}} \;\;\;\;\;\; (e_{1}^{x} = e^{x})
   \label{eq5}
\end{equation}

\noindent if $1 + (1 - q) x > 0$, and zero otherwise.  This reduces to the usual exponential function when $q = 1$, but when $q \ne 0$ it asymptotically approaches a power law in the limit $x \to \infty$.  When $q > 1$, the case of interest here, it asymptotically decays to zero.  The factor $p_0$ coincides with $p(0)$ if and only if $\delta=0$; $\kappa$ is a characteristic degree number.
The $q$-exponential function arises naturally as the solution of the equation ${dx}/{dt} = x^q$, which occurs as the leading behavior at some critical points.  It has also been shown \cite{bukman} to arise as the stationary solution of a nonlinear Fokker-Planck equation also known as the {\it Porous Medium Equation}. Various mesoscopic mechanisms (involving multiplicative noise) have already been identified  which yield this type of solution \cite{anteneodo}.

Finally, the $q$-exponential distribution also emerges from maximizing the entropy $S_q$ \cite{Tsallis} under a constraint that characterizes the number of degrees per node of the distribution. Let us briefly recall this derivation.
Consider the entropy
\begin{eqnarray}
S_q & \equiv & \frac{1-\int_0^\infty dk \, [p(k)]^q}{q-1}\\
\Bigl[S_1=S_{BG} & \equiv & -\int_0^\infty dk \, p(k) \ln p(k)\Bigr] \,, \nonumber
\end{eqnarray}

\noindent where we assume $k$ as a continuous variable for simplicity, and $BG$ stands for {\it Boltzmann-Gibbs}. If we extremize $S_q$ with
the constraints \cite{Tsallis}
\begin{equation}
\int_0^\infty dk \, p(k)=1
\label{constr1}
\end{equation}
and
\begin{equation}
 \frac{\int_0^\infty dk \, k \, [p(k)]^q}{ \int_0^\infty dk \,  [p(k)]^q  } = K >0\,,
 \label{constr2}
\end{equation}
we obtain
\begin{equation}
p(k)= \frac{e_q^{-\beta k}}{\int_0^\infty dk^\prime  \, e_q^{-\beta k^\prime}} \propto e_q^{- k/\kappa} \;\;\;\;(k \ge 0) \,,
\end{equation}
where the Lagrange parameter $\beta \equiv 1/\kappa$ is determined through Eq. (\ref{constr2}). Both
constraints (\ref{constr1}) and (\ref{constr2}) impose $q<2$.

Now to arrive at the Ansatz (\ref{eq4}) that we have used in this paper, we must provide some plausibility to the factor $k^{\,\delta}$ in front of the $q$-exponential. It happens this factor is the most frequent form of density of states in condensed matter physics (it exactly corresponds to systems of arbitrary dimensionality whose quantum energy spectrum is proportional to an arbitrary power of the wave-vector of the particles or quasi-particles; depending on the system, $\delta$ can be positive, negative, or zero, in which case the Ansatz reproduces a simple $q$-exponential). Such density of states concurrently multiplies the Boltzmann-Gibbs factor, which is here naturally represented by $e_q^{- k/\kappa}$.  In addition to this, Ansatz (\ref{eq4}) provided very satisfactory results in financial models where a plausible scale-free network basis was given to account for the distribution of stock trading volumes \cite{Osorio}.
An interesting financial mechanism using multiplicative noise has been recently proposed \cite{Queiros} which precisely leads to a stationary state distribution of the form (\ref{eq4}). It is for this ensemble of heuristic reasons that we checked the form (4). The numerical results that we obtained proved a posteriori that this choice was a good one.

\begin{figure*}[!hbp]
\begin{center}
\setlength{\unitlength}{0.6pt}
   \begin{picture}(700,700)
   \put(0,0){\includegraphics[width=75mm]{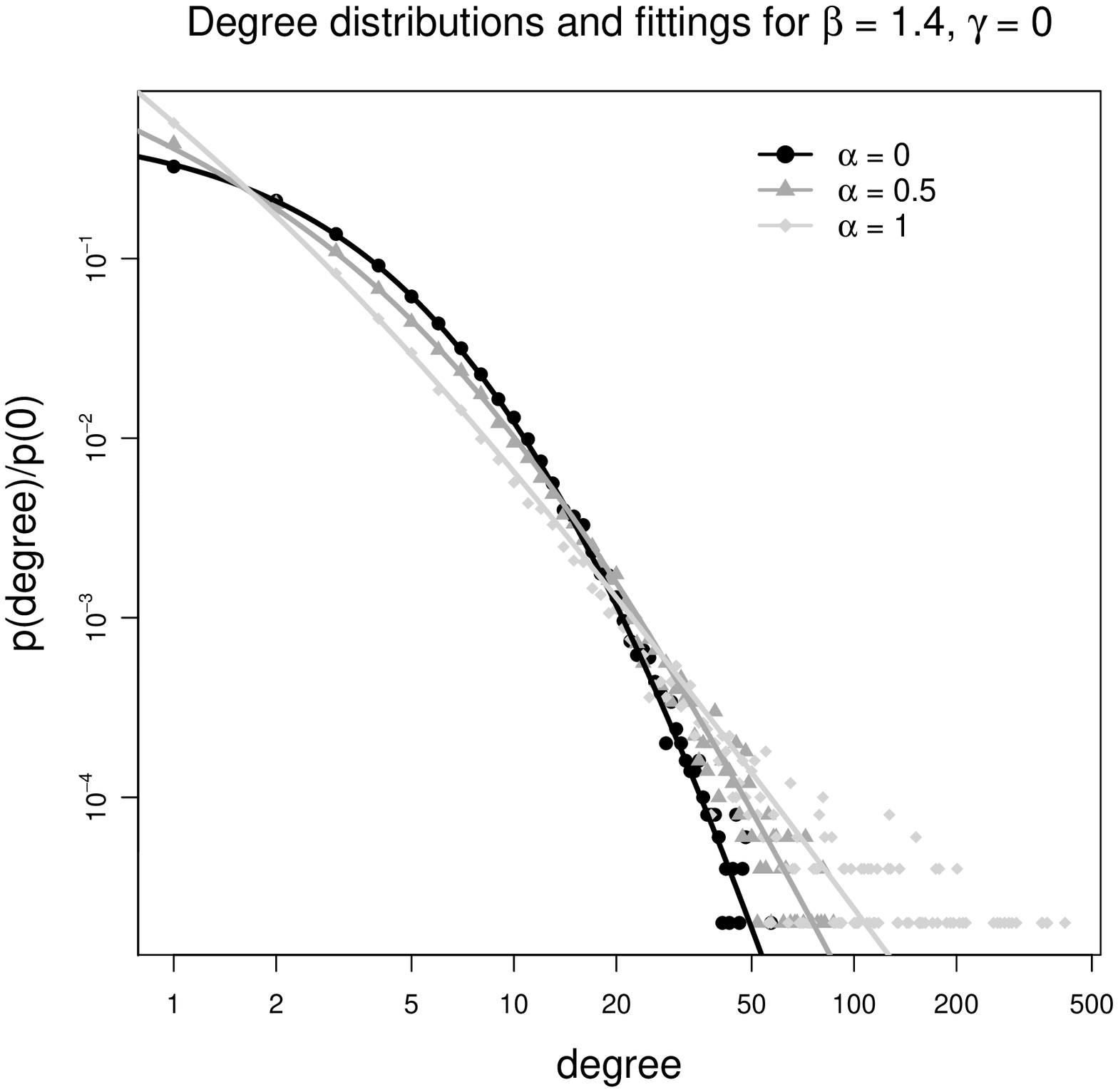}}
   \put(0,0){\small (c)}
   \put(350,0){\includegraphics[width=75mm]{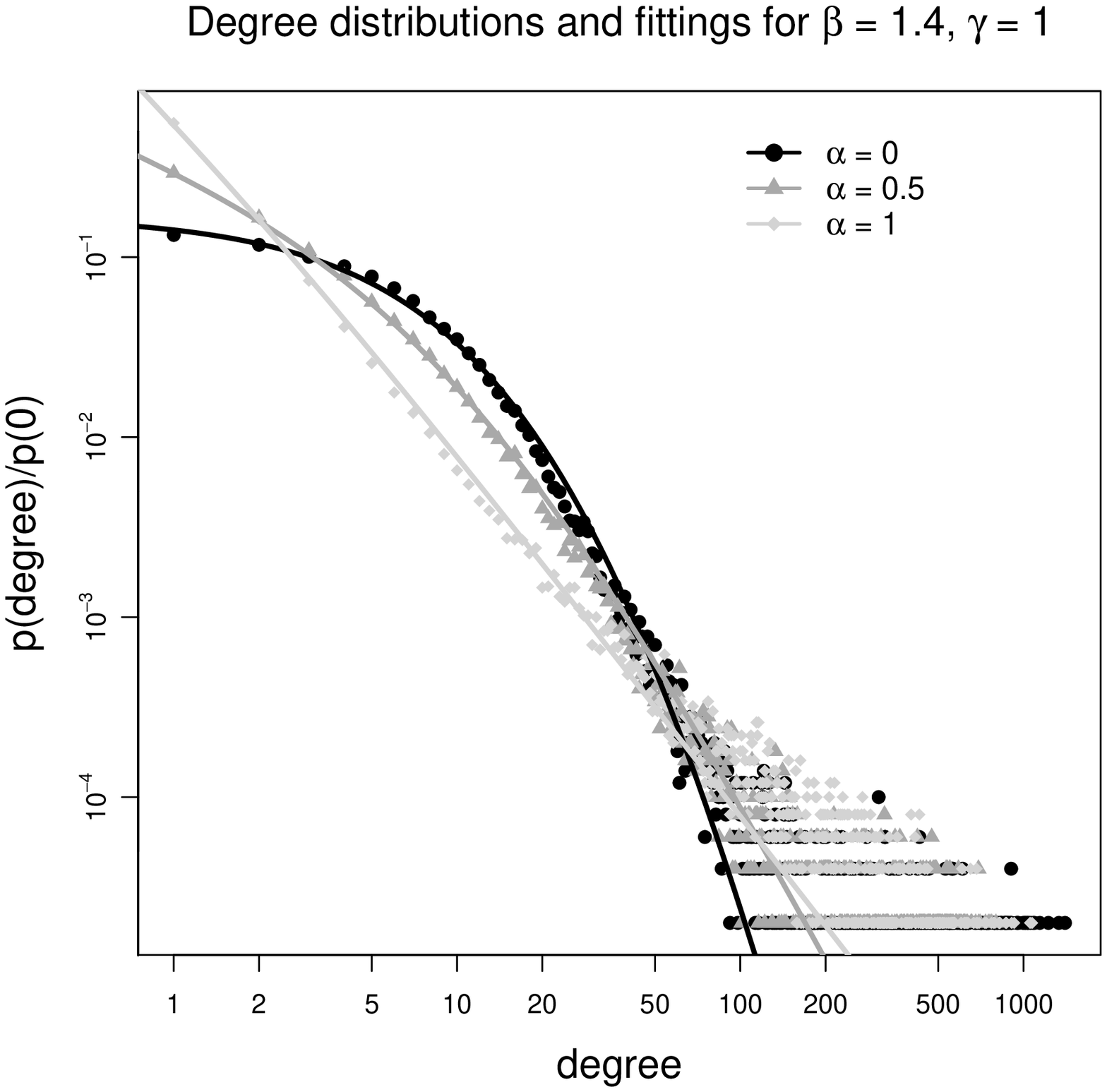}}
   \put(350,0){\small (d)}
   \put(0,350){\includegraphics[width=75mm]{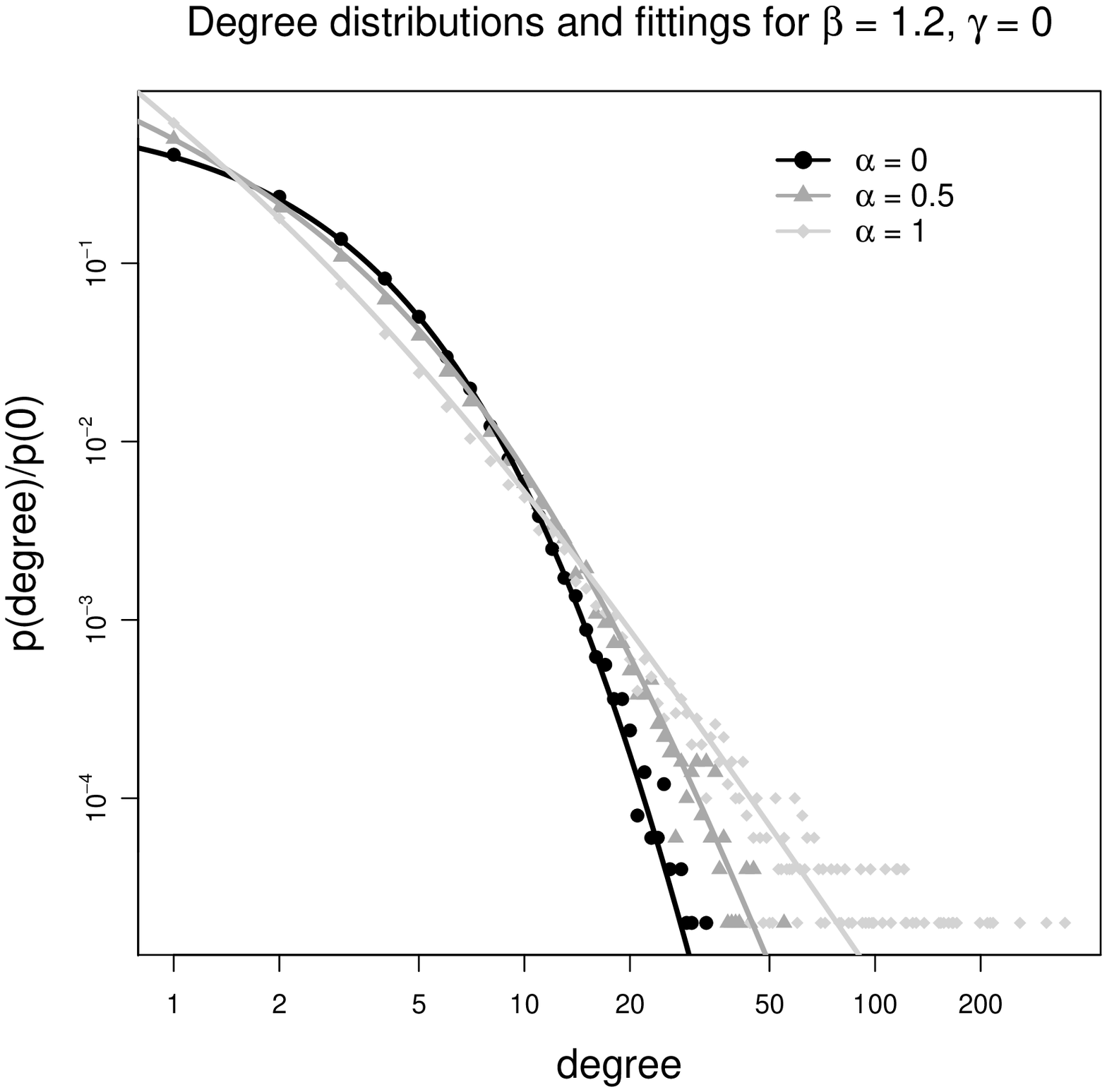}}
   \put(0,350){\small (a)}
   \put(350,350){\includegraphics[width=75mm]{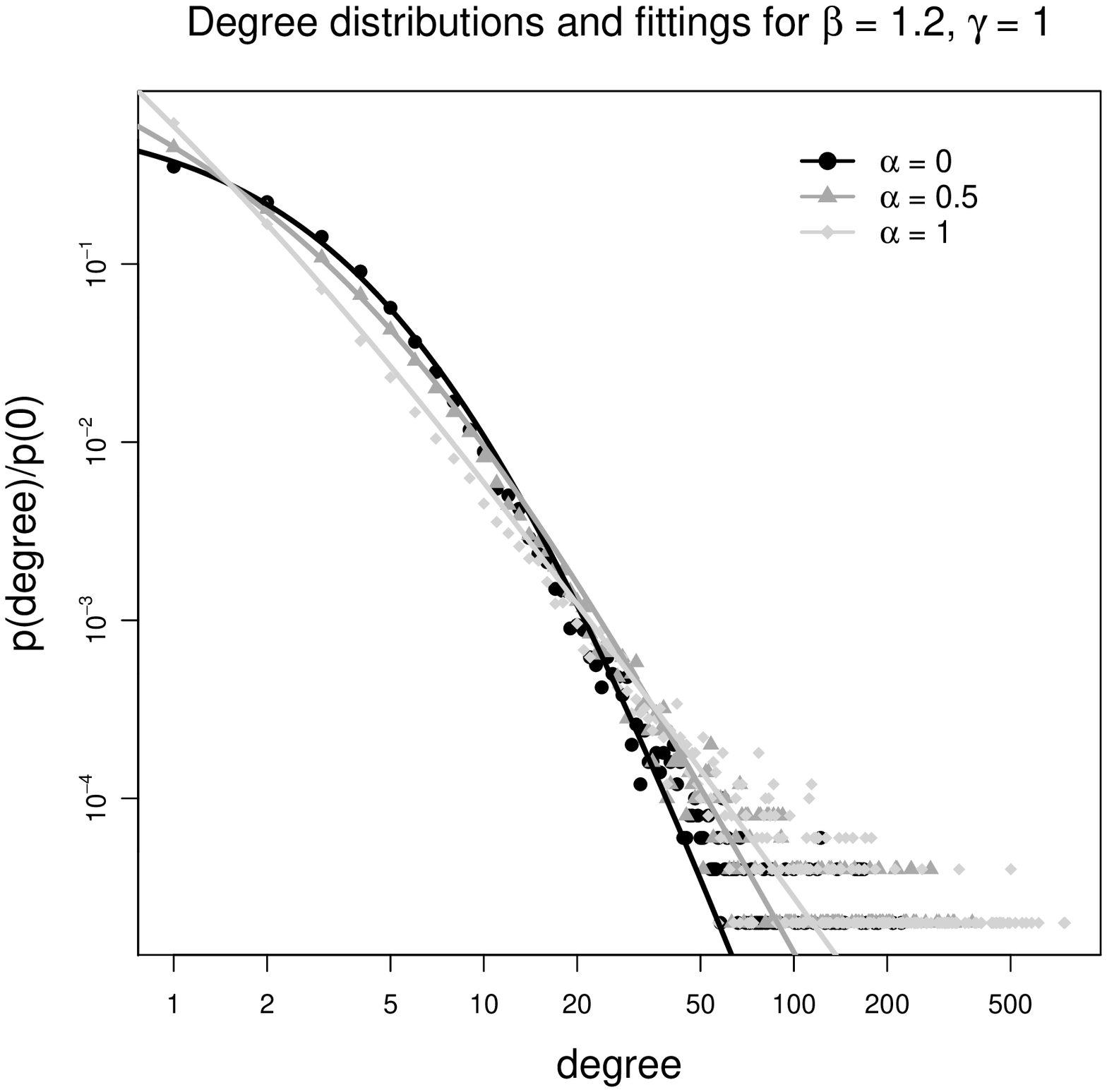}}
   \put(350,350){\small (b)}
   \end{picture}
\end{center}
\caption{Degree distributions and fits to a $q$-exponential for simulations of networks with $N=5000$ and $10$ realizations.  The dots correspond to the empirically observed frequency of each degree; the lowest row of dots in each case corresponds to observing one node with that degree.  The solid curves represent the best fit to a $q$-exponential.  In each case $\alpha$ has the three values $\{0, 0.5, 1\}$, corresponding to black, blue and red respectively.  (a) $\beta = 1.2, \gamma = 0$; (b) $\beta = 1.2, \gamma = 1$; (c) $\beta = 1.4, \gamma = 0$ and (d) $\beta = 1.4, \gamma = 1$.  Note that the scale of the  x-axes changes.   The parameters of the fitted generalized $q$-exponential functions are given in Table \ref{fitted values}.}
\label{degree distributions}
\end{figure*}

\begin{table*}[!htp]
  \caption{Parameters for the best fit to a $q$-exponential function for networks with different parameters. The first three columns are the parameters of the network model, and the next three columns are the parameters for the fit to the $q$-exponential.  The exponent $b$ is defined by $b \equiv \frac{1}{q-1} -\delta$ (see the text). The last two columns are $p$-values for nonparametric statistical Kolmogorov-Smirnov (K-S) and Wilcoxon rank sum (W) tests. The standard acceptance criterium is to have $p >0.05$, i.e., less than one failure in twenty. The asterisk depicts the one case where the null hypothesis was rejected. Consequently, if we demand that {\it both} K-S and W tests are satisfied, we obtained failure in only one among the twelve cases that we have analyzed.}
   \begin{ruledtabular}
    \begin{tabular}{ccccccccc}
     \multicolumn{3} {c}{\bf Network model} & \multicolumn{3} {c}{\bf Fitted parameters} & & \multicolumn{2} {c}{\bf $p$-values for nonparametric tests} \\ 
     $ \alpha $ & $\beta$ & $\gamma$ & $q $   & $\kappa$ & $\delta$ & $b$ & K-S test & W test \\ \hline 
         0  & 1.2   & 0      & 1.08 & 1.7    & 0                & 12.5  & 0.90            & 0.54 \\ 
        0.5 & 1.2   & 0      & 1.2  & 2.1    & -0.6         & 5.6   & 0.91            & 0.50 \\ 
         1  & 1.2   & 0      & 1.65 & 2.75   & -1.4         & 2.94  & 1.0             & 0.80 \\ 
         0  & 1.2   & 1      & 1.21 & 1.5    & 0            & 4.76  & 0.80          & 0.429 \\ 
        0.5 & 1.2   & 1      & 1.38 & 1.8    & -0.6         & 3.23  & 0.15          & 0.096 \\ 
         1  & 1.2   & 1      & 2.1  & 2.8    & -1.5         & 2.41  & 0.76          & 0.65 \\ 
         0  & 1.4   & 0      & 1.16 & 1.91   & 0                & 6.25  & 1.0             & 0.83 \\ 
        0.5 & 1.4   & 0      & 1.31 & 2.35   & -0.6         & 3.83  & 1.0             & 0.95 \\ 
         1  & 1.4   & 0      & 1.85 & 3.2    & -1.4         & 2.58  & 0.07              & 0.03*\\ 
         0  & 1.4   & 1      & 1.16 & 5.4    & 0            & 6.25  & 0.96            & 0.92 \\ 
        0.5 & 1.4   & 1      & 1.42 & 4.5    & -0.6             & 2.98  & 0.73          & 0.44 \\ 
         1  & 1.4   & 1      & 2.9  & 3      & -1.5         & 2.03  & 0.24            & 0.35 \\ 
    \end{tabular}
    \end{ruledtabular}
\label{fitted values}
\end{table*}

To study the node degree distribution $p(k)$, i.e. the frequency with which nodes have $k$ neighbors, we simulate 10 realizations of networks with $N=5000$ for different values of the parameters $\alpha, \beta$ and $\gamma$.  Some results are shown in Figure \ref{degree distributions}. We fit $q$-exponential functions to the empirical distributions using the Gauss-Newton algorithm for nonlinear least-squares estimates of the parameters. Due to limitations of the fitting software we used, we had to manually correct the fitting for the tail regions of the distribution. In Table \ref{fitted values} we give the parameters of the best fits for various values of $\alpha$, $\beta$, and $\gamma$, demonstrating that the degree distribution depends on all three parameters. The solid curves in Figure \ref{degree distributions} represent the best fit to a $q$-exponential.

The fits to the $q$-exponential are extremely good in every case.  To test the goodness of fit, we performed Kolmogorov-Smirnov (KS) and Wilcoxian (W) rank sum tests.  Due to the fact that the $q$-exponential is defined only on $[0,\infty)$, we used a two sample K-S test \cite{Bickel Doksum}.  To deal with the problem that the data are very sparse in the tail, we excluded data points with sample probability less than $10^{-4}$.  For the K-S test the null hypothesis is never rejected, and for the W test one case out of twelve is rejected, with a $p$~value of $0.03$.  Thus we can conclude that there is no evidence that the $q$-exponential is not the correct functional form.

\begin{figure*}[!h]
\begin{center}
    \includegraphics[width=75mm]{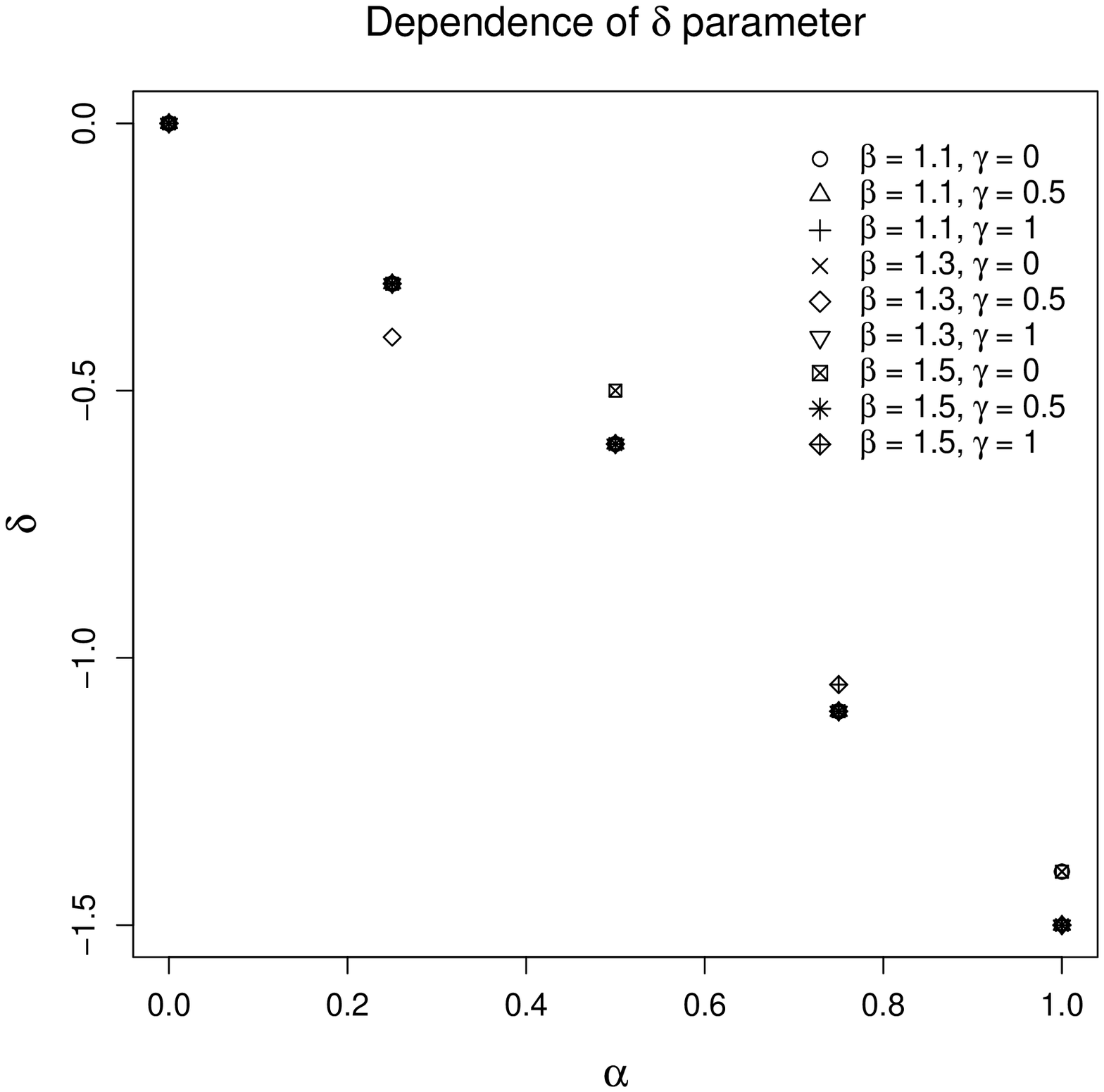}
\end{center}
\caption{Dependence of the q-exponential parameter $\delta$ on the network parameters $\alpha$, $\beta$, and $\gamma$.}
\label{param delta dependency}
\end{figure*}
\begin{figure*}[!hbp]
\begin{center}
\setlength{\unitlength}{0.6pt}
   \begin{picture}(700,700)
   \put(0,0){\includegraphics[width=50mm]{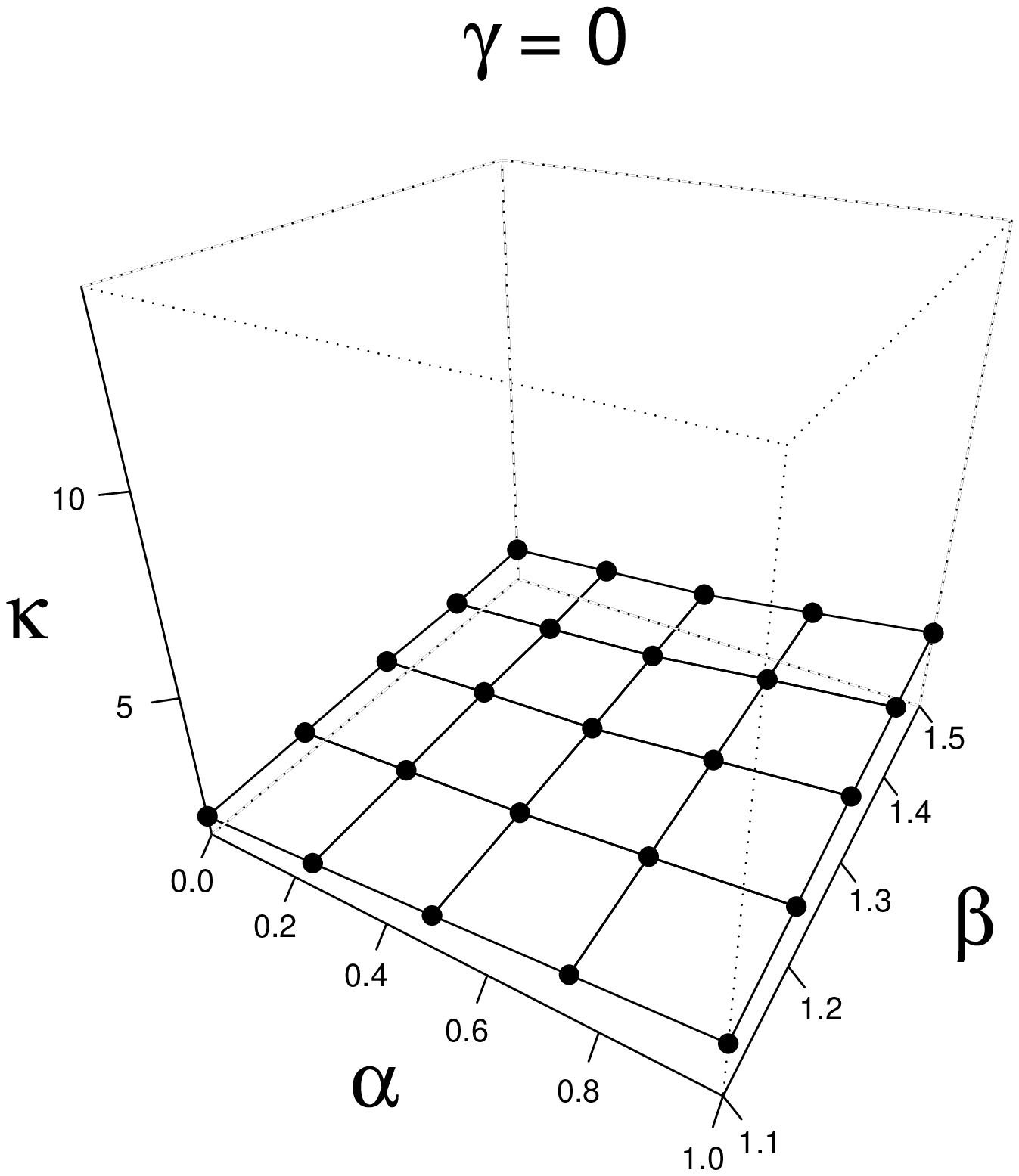}}
   \put(0,0){\small (d)}
   \put(233,0){\includegraphics[width=50mm]{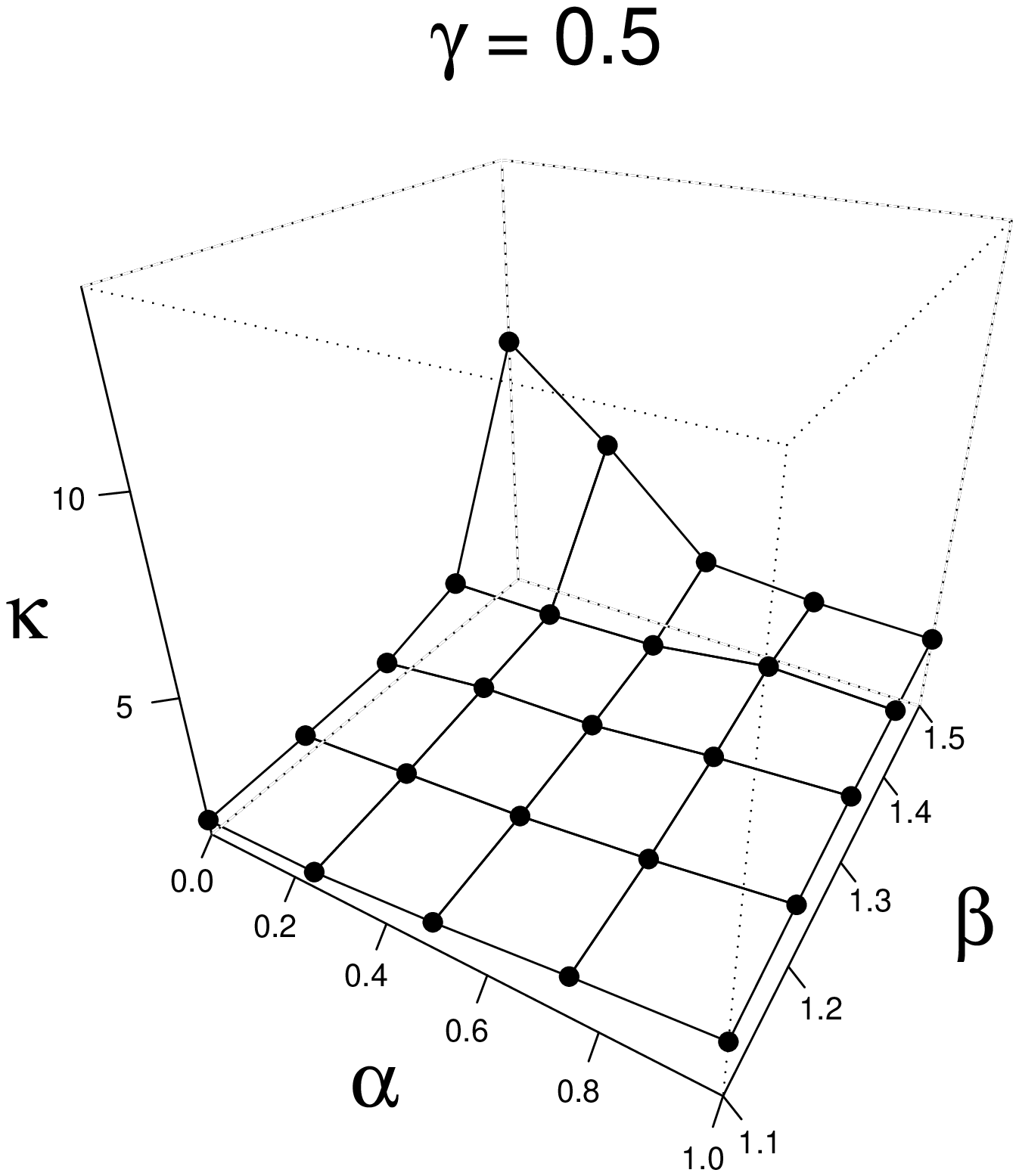}}
   \put(233,0){\small (e)}
   \put(466,0){\includegraphics[width=50mm]{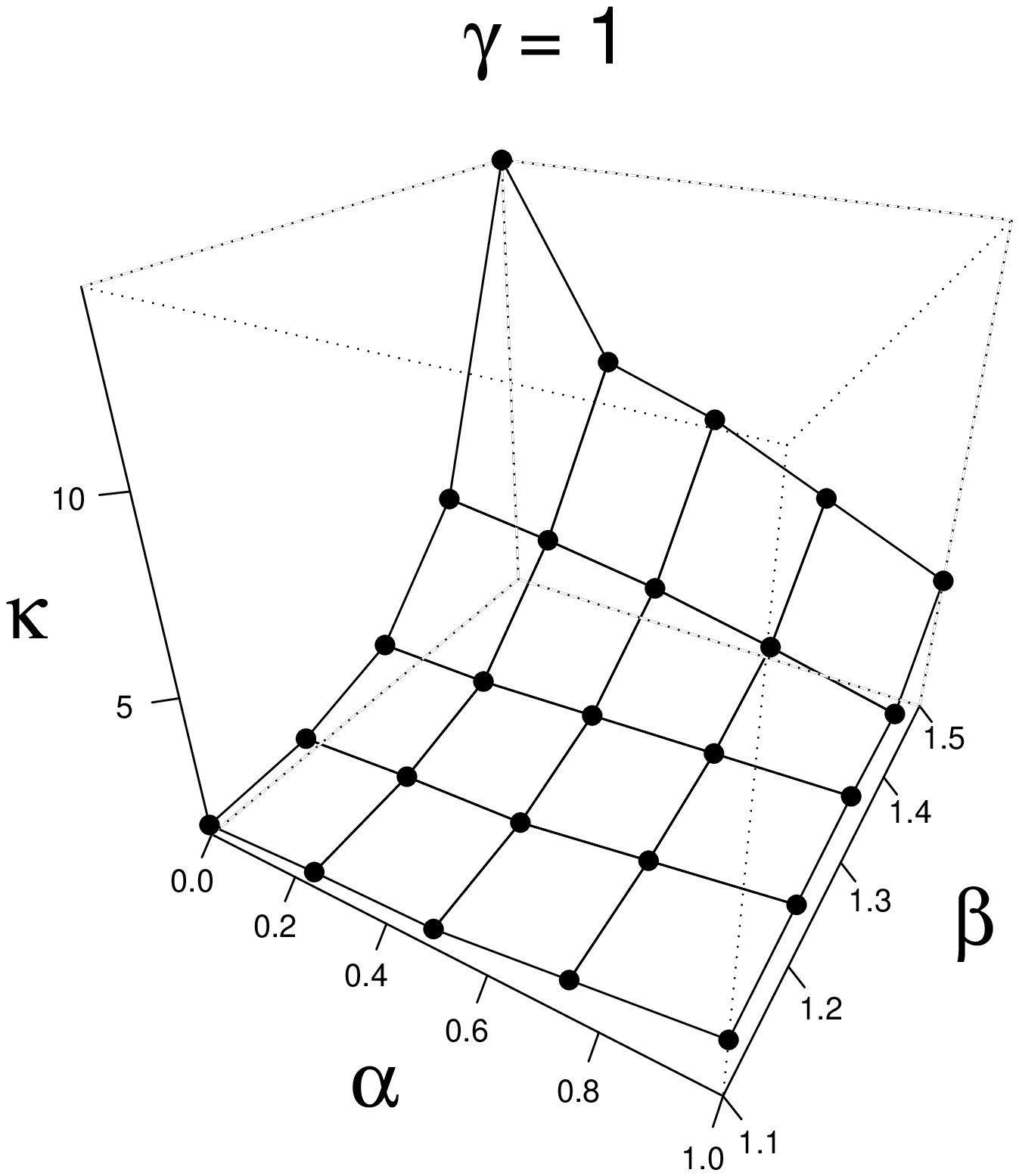}}
   \put(466,0){\small (f)}

   \put(0,300){\includegraphics[width=50mm]{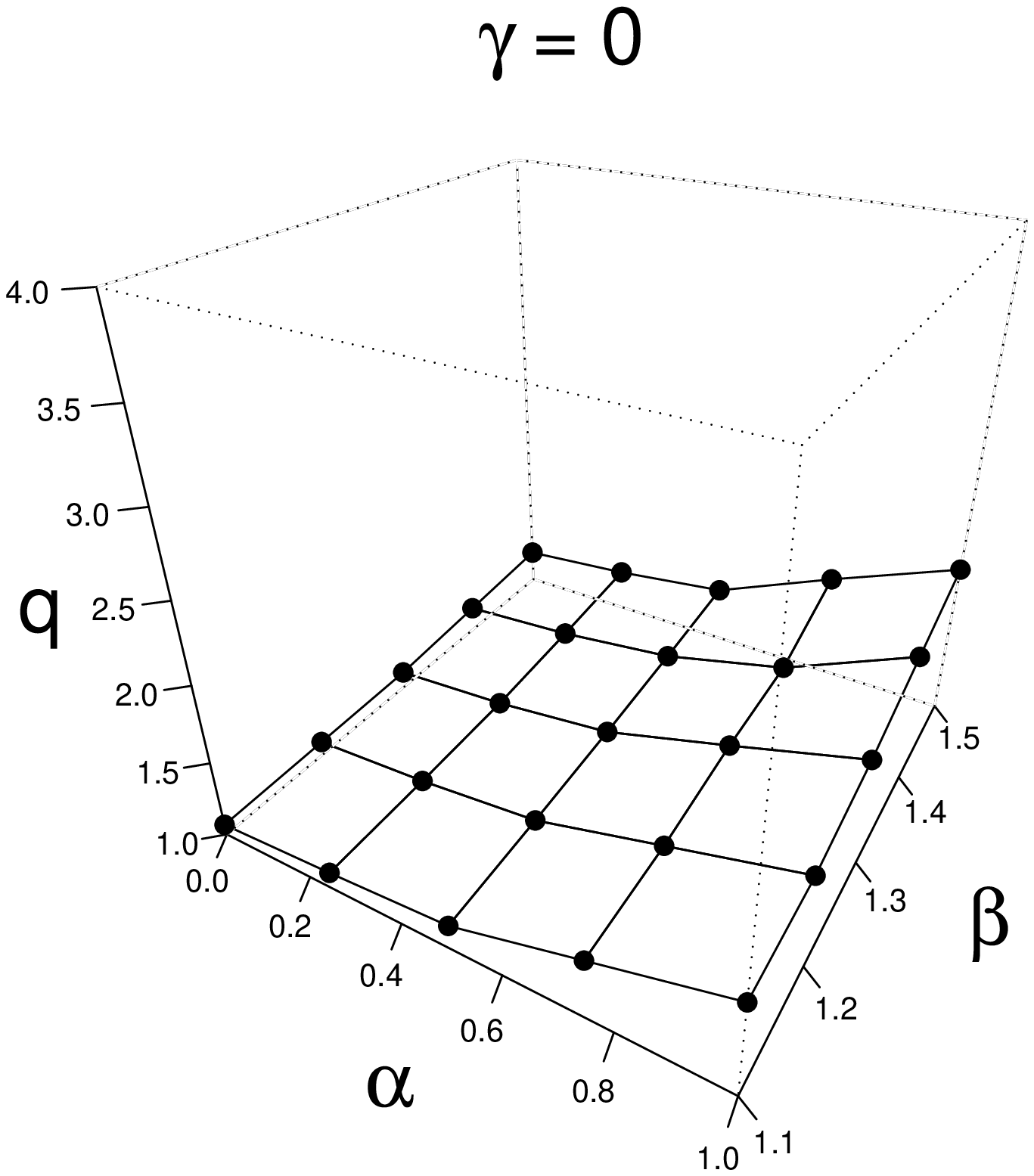}}
   \put(0,300){\small (a)}
   \put(233,300){\includegraphics[width=50mm]{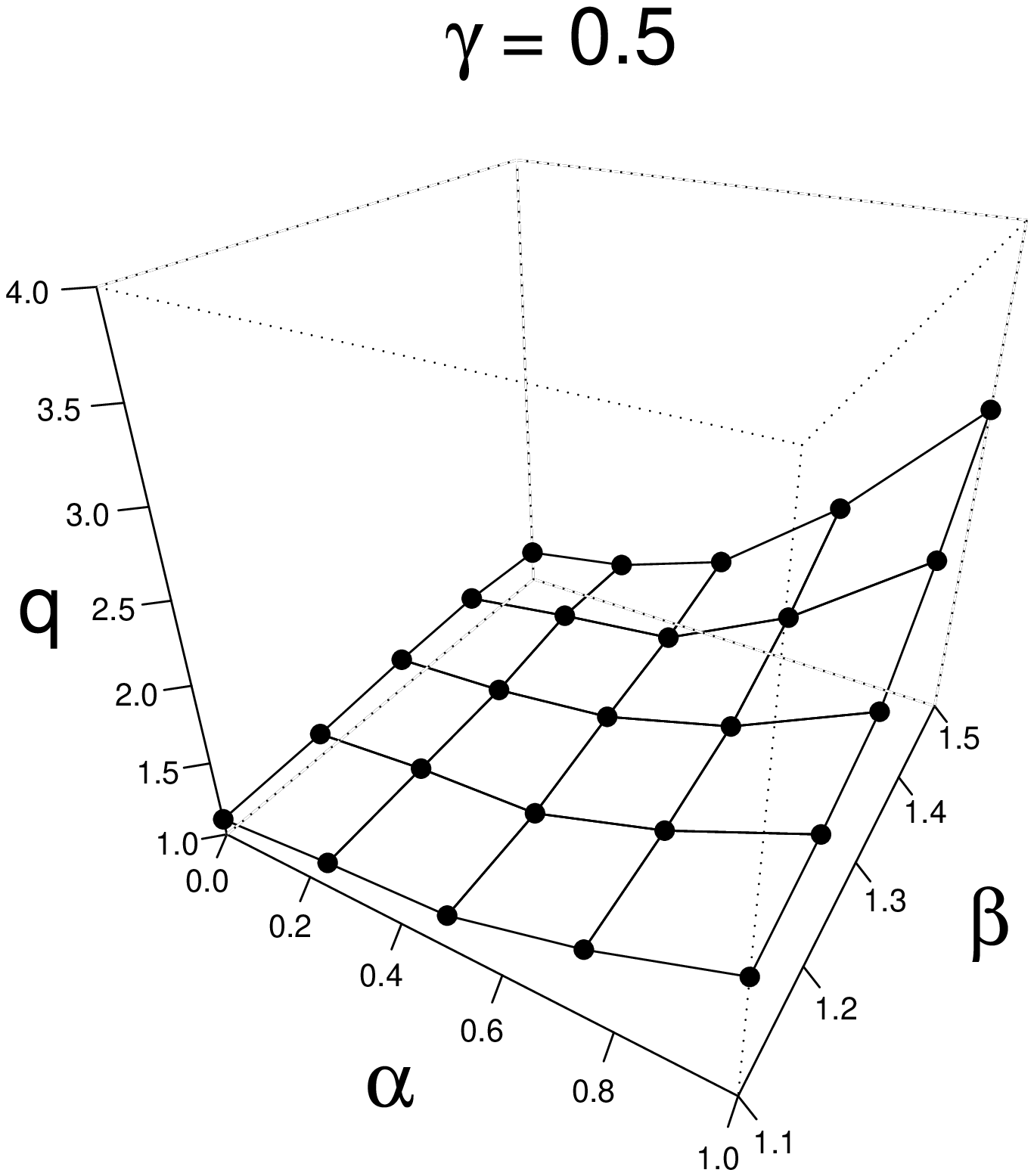}}
   \put(233,300){\small (b)}
   \put(466,300){\includegraphics[width=50mm]{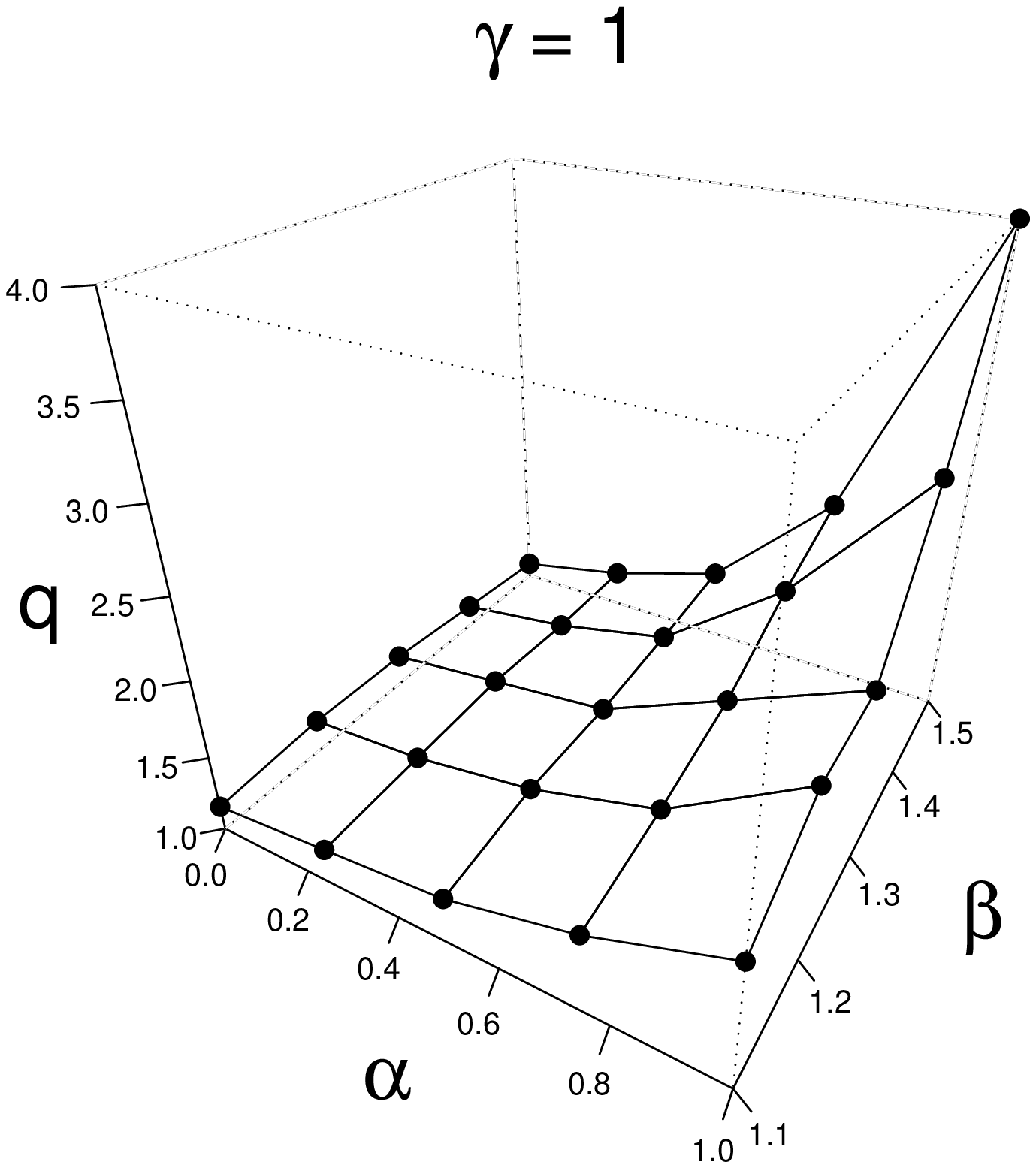}}
   \put(466,300){\small (c)}
   \end{picture}
\end{center}
\caption{Dependencies of q-exponential parameters $q$ and $\kappa$ that were fitted to network models.}
\label{param dependencies}
\end{figure*}
\begin{figure*}[!h]
\begin{center}
    \includegraphics[width=75mm]{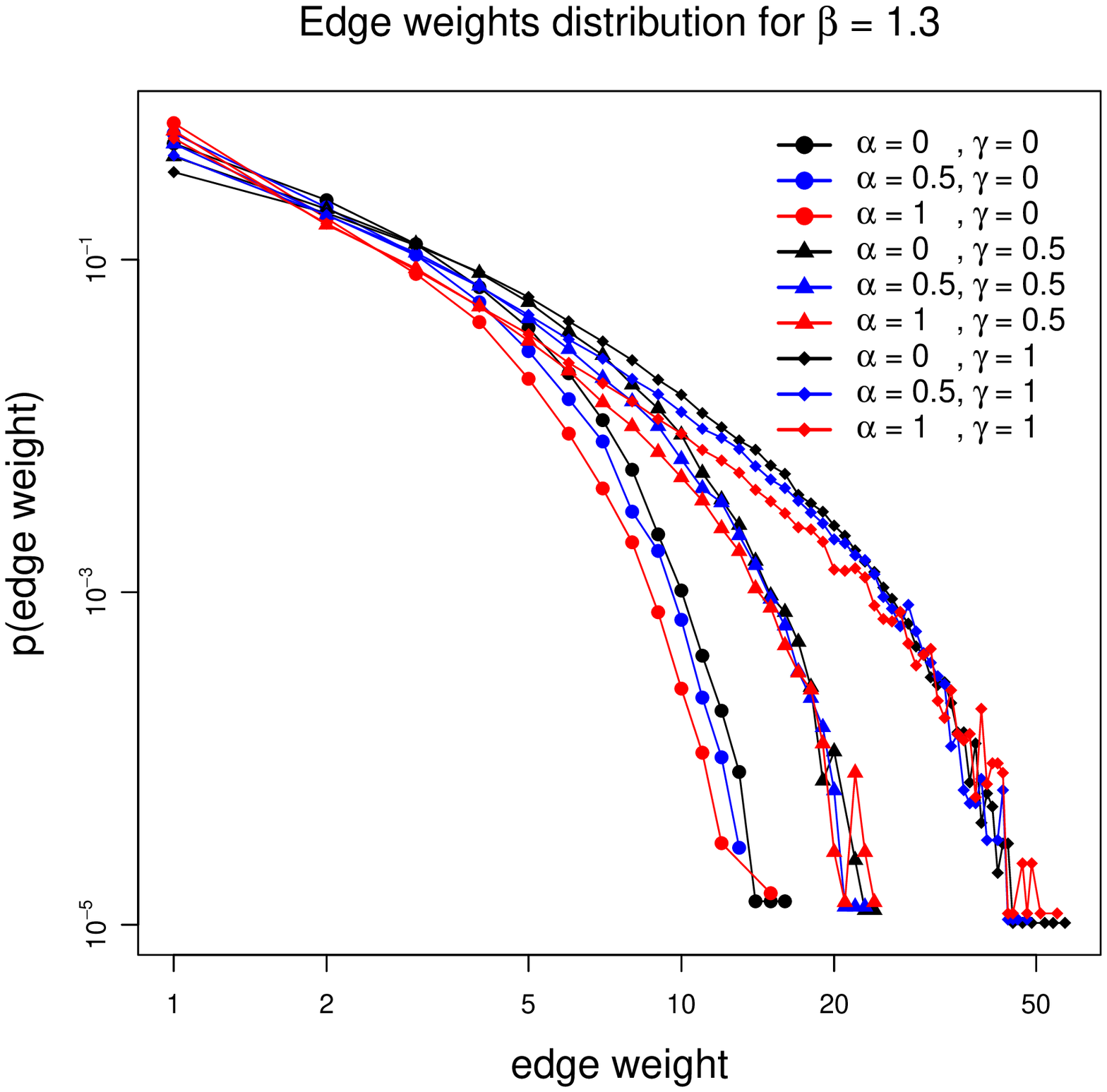}
\end{center}
\caption{Distribution of edge weights. Edge weights represent the number of successfully created feedback cycles in which an edge participated. The parameter $\beta = 1.3$, but $\alpha$ and $\gamma$ vary.  These calculations are based on 100 realizations of networks growing to $N=500$. The edge weights distribution experiences only a slight change to the right when increasing distance decay parameter $\beta$ while varying $\alpha$ but keeping $\gamma$ constant.
}
\label{edge distribution}
\end{figure*}

From Eq. (4) we straightforwardly verify that, in the $k\to\infty$ limit, we obtain (see also Figure \ref{degree distributions}) a Pareto distribution, of the form $ak^{-b}$, where $a \equiv p_0(\frac{\kappa}{q-1})^{1/(q-1)}$ and $b\equiv \frac{1}{q-1} -\delta$. This corresponds to scale-free behavior, i.e. the distribution remains invariant under the scale transformation $k \to Kk$.  In general, however, scale free behavior is only approached asymptotically, and the $q$-generalized exponential distribution, which contains the Pareto distribution as a special case, gives a much better fit.

\textbf{Parameters of model vs. $q$-exponential}. To understand how the parameters of the $q$-exponential depend on those of the model, we estimated the parameters of the $q$-exponential for $\alpha = \{0, 0.25, 0.5, 0.75, 1\}$, $\beta = \{1.1, 1.2, 1.3, 1.4, 1.5\}$ and $\gamma = \{0, 0.5, 1\}$.  Figure~\ref{param delta dependency} studies the dependence of $\delta$ on the graph parameters, and Figure~\ref{param dependencies} studies the dependence of $q$ and $\kappa$.

It is clear that $\delta$ depends solely on the attachment parameter $\alpha$. The other two q-exponential parameters ($q$ and $\kappa$) depend on all three model parameters. The parameter $\kappa$ diverges when $\beta$ and $\gamma$ grow large and $\alpha = 0$. The $q$ parameter grows rapidly as each of the three model parameters increase.

In Figure \ref{edge distribution} we study the distribution of edge weights, where an edge weight is defined as the number of times an edge participates in the construction of a feedback cyle (i.e. how many times it is traversed during the search leading to the creation of the cycle).  From this figure it is clear that this property is nearly independent of the attachment parameter $\alpha$, but is strongly depends on the routing parameter $\gamma$.

\section{CONCLUSIONS}

In this paper we have presented a generative model for creating graphs representing feedback networks. The construction algorithm is strictly local, in the sense that any given step in the construction of a network only requires information about the nearest neighbors of nodes.  Nonetheless, the resulting networks display long-range correlations in their structure.  This is reflected in the fact that the $q$-exponential distribution, which is associated with long-range correlation in problems in statistical mechanics, provides a good fit to the degree distribution.

We think this adds an important contribution to the literature on the generation of networks by illustrating a mechanism that specifically focuses on the competition between consolidation by adding cycles, which represent stronger feedback within the network, and growth in size by simply adding more nodes.  
In future work, we hope to apply the present model to real networks such as biotech intercorporate networks, medieval trading networks, marriage networks, and other real examples.

\begin{acknowledgments}
Partial sponsoring from SI International and AFRL is acknowledged.
\end{acknowledgments}

\end{document}